\documentclass[cameraready]{Interspeech}
\title{Prosodic Boundary-Aware Streaming Generation for LLM-Based TTS with Streaming Text Input}

\author[affiliation={1,*}]{Changsong}{Liu}
\author[affiliation={2,*}]{Tianrui}{Wang}
\author[affiliation={3}]{Ye}{Ni}
\author[affiliation={1}]{Yizhou}{Peng}
\author[affiliation={1}]{Eng Siong}{Chng}

\address{
    $^1$ Nanyang Technological University, Singapore \\
    $^2$ Tianjin University, China \\
    $^3$ Southeast University, China
}

\email{changsong.liu@ntu.edu.sg, wangtianrui@tju.edu.cn, niye@seu.edu.cn, peng.yizhou@ntu.edu.sg, ASESChng@ntu.edu.sg}


\usepackage{comment}

\usepackage{amsmath}
\usepackage{multirow}
\usepackage{booktabs}
\usepackage{hyperref}

\begin{document}

\maketitle
\begingroup
\renewcommand\thefootnote{*}
\footnotetext{These authors contributed equally to this work.}
\endgroup

\begin{abstract}

Streaming TTS that receives streaming text is essential for interactive systems, yet this scheme faces two major challenges: unnatural prosody due to missing lookahead and long-form collapse due to unbounded context. We propose a prosodic-boundary-aware post-training strategy, adapting a pretrained LLM-based TTS model using weakly time-aligned data. Specifically, the model is adapted to learn early stopping at specified content boundaries when provided with limited future text. During inference, a sliding-window prompt carries forward previous text and speech tokens, ensuring bounded context and seamless concatenation. Evaluations show our method outperforms CosyVoice-Style interleaved baseline in both short and long-form scenarios. In long-text synthesis, especially, it achieves a 66.2\% absolute reduction in word error rate (from 71.0\% to 4.8\%) and increases speaker and emotion similarity by 16.1\% and 1.5\% relatively, offering a robust solution for streaming TTS with incremental text.

\end{abstract}


\section{Introduction}

Streaming text-to-speech (TTS) with streaming text input aims to generate speech in real-time as text arrives, which is in high demand for applications such as dialogue systems and speech-to-speech translation \cite{dang2024zero, moshi}.

The usability of such systems is largely determined by synthesis latency, which mainly arises from two sources: the waiting time for text segment accumulation, and the model's inference time for converting text into audio. While the latter can be effectively mitigated by optimizing the model size or introducing causal structures \cite{pamisetty2025stream, sun2025zero}, achieving ultra-low latency requires keeping the text accumulation window extremely small. This leads to the first core challenge: natural and high-quality speech synthesis heavily relies on sufficient contextual information. The model requires not only historical text to maintain coherence but also future text (lookahead) to accurately predict prosodic features such as stress and pauses \cite{sheng2025syncspeechlowlatencyefficientdualstream}. Thus, a restricted receptive field results in unnatural prosody. 
Existing approaches \cite{dang2024zero, dekel2024speak, dang2024livespeech} attempt to address this using local context modeling, but typically require complex causal modifications to the attention mechanism and rely on precise text–speech forced alignment \cite{bai20223, kim2023transduce}.

\begin{figure*}[!ht]
    \centering
    \includegraphics[width=0.98\textwidth]{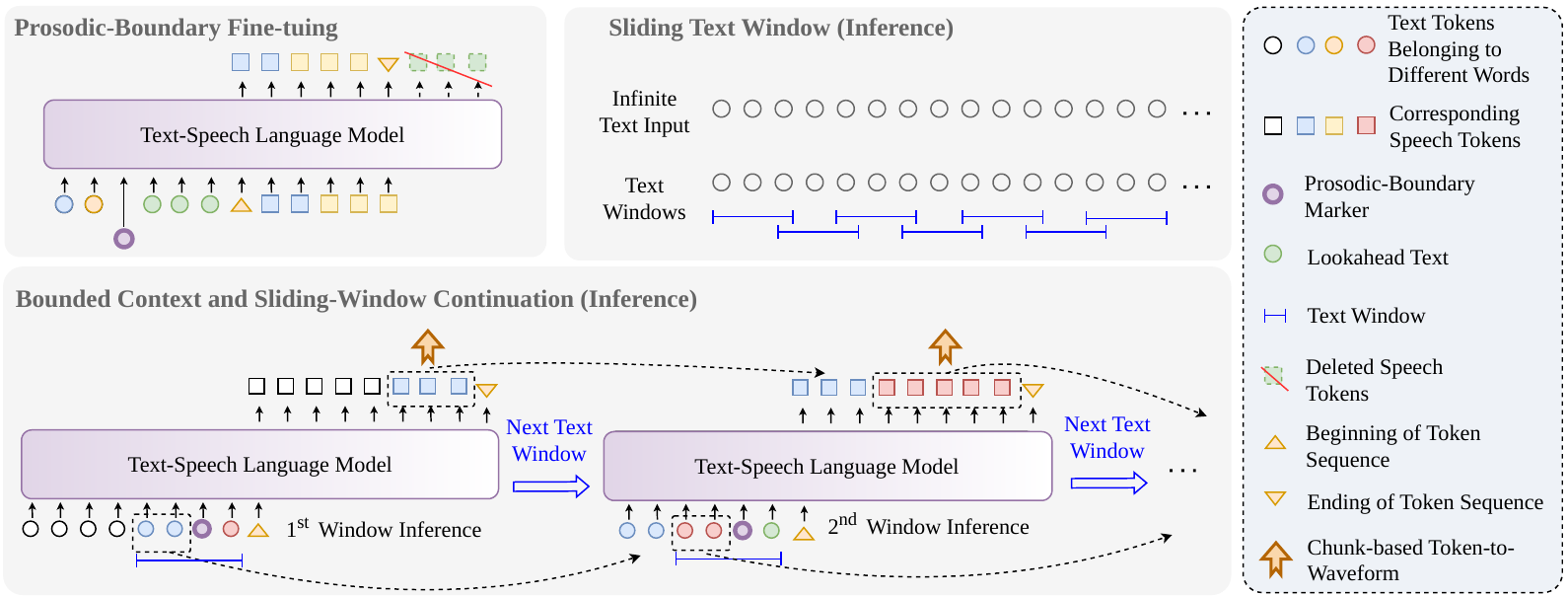}
    \vspace{-0.3cm}
    \caption{The proposed fine-tuning and inferencing pipeline for prosodic boundary-aware streaming LLM-based TTS generation.}
    \label{fig:diagram}
\end{figure*}

Meanwhile, with the widespread adoption of large language model (LLM) architectures in the speech domain, 
modern TTS systems increasingly adopt cross-modality modeling that interleaves text and speech tokens \cite{audiolm, valle, wescon, interleave1, wang2025streammel, yang2024interleaved}. Although such architectures achieve state-of-the-art synthesis quality, enabling streaming generation with streaming text input introduces a second challenge: long-form performance collapse caused by unbounded generation history.
For instance, Cosyvoice-series \cite{du2024cosyvoice2, du2025cosyvoice} employs an interleaved arrangement of fixed numbers of text and speech tokens. Similarly, the streaming version of Qwen3-TTS \cite{hu2026qwen3} adopts a comparable interleaved structure, albeit arranging text and speech tokens in parallel. Yet in real-world scenarios with long-term continuous input, since the speech length corresponding to a single text token varies enormously, the physical distance between text and its associated speech tokens gradually widens. This ultimately causes generation failure, making it difficult to support long-term streaming interactions. SpeakStream \cite{bai2025speakstreamstreamingtexttospeechinterleaved} alleviates this issue through character-level alignment and history truncation, but it relies heavily on precise alignment annotations.
These limitations motivate the following research question: \textbf{Can robust long-form streaming be achieved in LLM-based TTS with streaming text input using only weakly time-aligned data without architectural modifications?}


To tackle this question, we propose a novel post-training strategy for prosodic-boundary-aware adaptation that adapts existing LLM-based TTS models for robust streaming using only weakly time-aligned data. Our contributions are as follows:

\begin{itemize}
    \item We introduce a prosodic-boundary-aware adaptation combined with a windowed lookahead mechanism, allowing models to anticipate future text to improve prosody without requiring complex causal modifications.
    \item We design an acoustic prompting method utilizing the previous chunk's audio tail, which ensures seamless concatenation and mitigates generation collapse in long-form cross-modality continuous streaming.
    \item We demonstrate state-of-the-art streaming stability and robustness using only weakly time-aligned open-source data, significantly outperforming existing interleaved baselines in real-time deployment. Audio demos are available at:\footnote{\url{https://anonymous-demo-168.github.io/Prosodic-Boundary-Aware-Streaming-Text-TTS-demo/}}
\end{itemize}

\section{Methodology}
\label{sec:method}

\subsection{Prosodic-Boundary Marker}
To enable streaming generation while preserving prosodic naturalness, we formulate the input as a bifurcated sequence via a prosodic-boundary-aware marker, {$\texttt{marker}_{\texttt{boundary}}$}, to decouple the acoustic generation span from the broader prosodic context. As illustrated in Figure \ref{fig:diagram}, the model learns to treat this marker as a soft boundary. During inference, the marker is inserted every $k$ words, allowing the model to leverage limited future context for prosodic planning while preventing unbounded generation context.

\subsection{Training with Weakly Time-Aligned Supervision}

The model is adapted using word-level timestamps obtained from an off-the-shelf aligner, WhisperX \cite{bain2022whisperx}, to approximate candidate prosodic boundaries without manual annotation. Given an utterance with text tokens ${x}_{1:T}$, speech tokens ${s}_{1:L}$, and word-level boundaries (where word $j$ maps to a text-token span $[b_j, e_j]$ and audio-end time $a_j$), we apply \textbf{Dynamic Boundary Insertion}. During training, we stochastically decide with probability $p_\text{full}$ to use the full, unmodified utterance to preserve global coherence. Otherwise, we randomly sample a word index $m$ and insert the boundary marker into the text sequence:
\begin{equation}
 x' = (x_{1:e_m},\,\textbf{$\texttt{marker}_{\texttt{boundary}}$},\,{x}_{e_m+1:T}).
\label{eq:text_insert}
\end{equation}
The target speech sequence is truncated to the aligned audio position of word $m$. Let $r_s$ denote the speech-token frame rate and $\ell{\min}$ a minimum length. The truncated target length $L'$ is:
\begin{equation}
L' = \max\bigl(\ell_{\min},\,\lfloor a_m \cdot r_s \rfloor\bigr).
\label{eq:speech_trunc}
\end{equation}
The language model is then fine-tuned to predict the truncated stream ${s}_{1:L'}$ conditioned on $x'$. This procedure trains the model to interpret the marker as both a segmentation cue and a prosodic anchor, ensuring acoustic output is aligned only with the segment preceding the marker.

\subsection{Bounded Context and Sliding-Window Continuation}
During inference, input text is processed in chunks of $k$ words, with a lookahead of $f$ future words. For chunk index $t$, let ${x}_{t_{\text{cur}}}$ denote the current segment and ${x}_{t_{\text{fut}}}$ the lookahead text. The input sequence is constructed as:
\begin{equation}
{X}_t = \bigl({x}_{t_{\text{cur}}},\, \textbf{$\texttt{marker}_{\texttt{boundary}}$},\, {x}_{t_{\text{fut}}}\bigr).
\label{eq:infer_input}
\end{equation}
To maintain cross-chunk continuity, we employ a \textbf{Sliding-Window Prompt}. The first chunk is conditioned on a reference utterance (${x}_{\text{ref}}$ ,(${s}_{\text{ref}}$) for zero-shot voice cloning. For subsequent chunks, the prompt (${p}^x_t$, ${p}^s_t$) is replaced with the text and speech tokens synthesized in the previous step:
\begin{equation}
\bigl({p}^x_t,\, {p}^s_t\bigr)
=
\begin{cases}
\bigl({x}_{\text{ref}},\, {s}_{\text{ref}}\bigr), & t = 1, \\[2pt]
\bigl(\hat{{x}}_{t-1},\, \hat{{s}}_{t-1}\bigr), & t \geq 2.
\end{cases}
\label{eq:prompt}
\end{equation}
where $\hat{{x}}_{t-1}$ denotes the previously generated text segment corresponding to the current chunk (excluding lookahead), and $\hat{{s}}_{t-1}$ is the corresponding synthesized speech. This design keeps the Key-Value (KV) cache bounded by $\mathcal{O}(k+f)$ regardless of total sequence length, preventing both latency growth and long-form instability. Finally, the generated speech tokens are passed to a streaming vocoder for incremental waveform synthesis, enabling seamless concatenation across chunks.

\begin{table}[b]
\caption{Evaluation on streaming efficiency. Chunk size $k=5$ words; lookahead $f=2$ words for our proposed method.}
\label{tab:latency}
\centering
\small
\begin{tabular}{lcc}
\toprule
\textbf{System} 
& \textbf{RTF}$\downarrow$ 
& \textbf{TTFA (ms)}$\downarrow$ \\
\midrule
Interleaved & 0.843 & \underline{1414} \\
Sliding-Window  & \textbf{0.718} & 2588 \\
Boundary-Aware & \underline{0.782} & \textbf{1296} \\
\bottomrule
\end{tabular}
\end{table}

\begin{table*}[t]
\caption{Objective quality evaluation on Seed-TTS-Eval.
Chunk size $k=5$ words; lookahead $f=2$ words for our proposed method.}
\label{tab:quality}
\centering
\small
\begin{tabular}{lcccccc}
\toprule
\multirow{2}{*}{\textbf{System}} 
& \multicolumn{2}{c}{\textbf{WER (\%)} $\downarrow$}
& \multicolumn{2}{c}{\textbf{SPK-SIM} $\uparrow$}
& \multicolumn{2}{c}{\textbf{EMO-SIM} $\uparrow$} \\
\cmidrule(lr){2-3}
\cmidrule(lr){4-5}
\cmidrule(lr){6-7}
& Standard & Long-form 
& Standard & Long-form 
& Standard & Long-form \\
\midrule
Interleaved
& 7.48 & 70.97 
& 0.53 & 0.56 
& 0.899 & 0.899 \\

Sliding-Window 
& 6.03 & 7.83 
& 0.57 & 0.22 
& 0.912 & 0.857 \\

Boundary-Aware (Ours) 
& \textbf{4.03} & \textbf{4.77} 
& \textbf{0.64} & \textbf{0.65} 
& \textbf{0.918} & \textbf{0.912} \\
\bottomrule
\end{tabular}
\end{table*}

\begin{table*}[t]
\setlength\tabcolsep{1pt}
\caption{Subjective MOS evaluation on Seed-TTS-Eval.
Higher scores indicate better perceptual quality.}
\label{tab:mos}
\centering
\small
\begin{tabular}{lcccccc}
\toprule
\multirow{2}{*}{\textbf{System}} 
& \multicolumn{2}{c}{\textbf{MOS} $\uparrow$}
& \multicolumn{2}{c}{\textbf{SMOS} $\uparrow$}
& \multicolumn{2}{c}{\textbf{EMOS} $\uparrow$} \\
\cmidrule(lr){2-3}
\cmidrule(lr){4-5}
\cmidrule(lr){6-7}
& Standard & Long-form 
& Standard & Long-form 
& Standard & Long-form \\
\midrule
Interleaved
& 3.99 $\pm$ 0.16 & 3.18 $\pm$ 0.23
& 4.05 $\pm$ 0.15 & 3.24 $\pm$ 0.22
& 4.03 $\pm$ 0.15 & 3.21 $\pm$ 0.18 \\

Sliding-Window
& 3.43 $\pm$ 0.20 & 1.60 $\pm$ 0.18
& 3.18 $\pm$ 0.20 & 1.68 $\pm$ 0.18
& 3.25 $\pm$ 0.18 & 1.67 $\pm$ 0.17 \\

Boundary-Aware (Ours)
& \textbf{4.28 $\pm$ 0.14} & \textbf{4.13 $\pm$ 0.13}
& \textbf{4.25 $\pm$ 0.13} & \textbf{4.24 $\pm$ 0.13}
& \textbf{4.38 $\pm$ 0.12} & \textbf{4.19 $\pm$ 0.13} \\
\bottomrule
\end{tabular}
\end{table*}

\section{Experiments}

\subsection{Training Datasets}
We train our model on the sampled English subset of CommonVoice 13.0~\cite{commonvoice}, containing approximately 930k utterances (about \texttt{\~{}}1,000 hours). To ensure zero-shot evaluation integrity, we remove any training utterances overlapping with the Seed-TTS-Eval test set~\cite{seedtts}.

Before training, we pre-extract speech tokens, mel-spectrograms, text tokens, and speaker embeddings. Word-level timestamps are obtained using WhisperX \cite{bain2022whisperx} to approximate candidate prosodic boundaries. During fine-tuning, the probability of using the full utterance is set to $p_\text{full} = 0.15$. The speech tokenizer operates at a frame rate of $r_s = 25\,$Hz, and the minimum truncated length is $\ell_{\min} = 5$ frames.

\subsection{Evaluation Datasets}

We evaluate the system under two complementary tiers to assess both sentence-level quality and long-form robustness.

\textbf{Standard-form Evaluation}: We use the Seed-TTS-Eval benchmark \cite{seedtts}, which consists of short read-style sentences with reference prompts. This set evaluates standard sentence-level streaming synthesis.

\textbf{LLM-expanded Long-form Evaluation}: To stress-test long-form stability, we construct an expanded benchmark by extending each Seed-TTS-Eval sentence into a coherent paragraph of 280--320 words using DeepSeek-V3 \cite{deepseekai2024deepseekv3technicalreport}, while preserving the original sentence as the opening segment. This setting evaluates prosodic consistency and synthesis stability during extended monologues.

\subsection{Evaluation Metrics}

Streaming efficiency is evaluated using Time-to-First-Audio (TTFA) and Real-Time Factor (RTF). TTFA measures the latency between the initial synthesis request and the first decodable audio chunk, reflecting perceived responsiveness. An RTF $< 1.0$ indicates real-time synthesis capability. To avoid hardware warm-up effects, we report averages over 50 trials following two warm-up runs for each utterance. The latency metrics are measured on a single NVIDIA A40 GPU (48\,GB VRAM).

For synthesis quality, we follow the Seed-TTS-Eval protocol \cite{seedtts}. Linguistic accuracy is measured by Word Error Rate (WER) using the Parakeet-TDT-0.6B-v2 model \cite{nvidia_parakeet_tdt_0_6b_v2_2026}, which supports long-form transcription beyond the typical 30-second limit. Acoustic fidelity is evaluated via cosine similarities of utterance embeddings using WavLM-Large for Speaker Similarity (SPK-SIM) and emotion2vec+ \cite{ma-etal-2024-emotion2vec} for Emotional Similarity (EMO-SIM). We further conduct subjective listening tests where 20 evaluators rate the generated speech on a 5-point Likert scale for Intelligibility (MOS), Speaker Similarity (SMOS), and Emotion Similarity (EMOS). For similarity ratings, generated samples are compared against the reference prompts. All subjective scores are reported as mean scores with 95\% confidence intervals computed via standard $t$-tests.

\subsection{Experimental Setup and Baselines}

\subsubsection{Base Architecture}
We adopt CosyVoice2 \cite{du2024cosyvoice2} as the foundation model. The architecture consists of a Qwen-based LLM that generates speech tokens, followed by a flow-matching module and a pretrained HiFi-GAN vocoder \cite{hifigan} for waveform synthesis. During training, only the LLM is fine-tuned using the proposed method (Section~\ref{sec:method}); the flow-matching and vocoder remain frozen.

\subsubsection{Baselines and Proposed Method}
\label{Baselines}
In this section, we describe the baselines. All systems use a fixed chunk size of $k=5$ words for streaming input.

\textbf{Interleaved Baseline}: We use the native streaming implementation of CosyVoice2 (\texttt{inference\_{bistream}}), where text and speech tokens are interleaved at a 5:15 ratio within a single sequence. The KV cache grows throughout generation, and streaming vocoding is used for incremental waveform generation. We use the original pre-trained weights without boundary-aware adaptation.

\begin{figure*}[!t]
    \centering
    \includegraphics[width=1.2\textwidth, height=0.28\textheight, keepaspectratio]{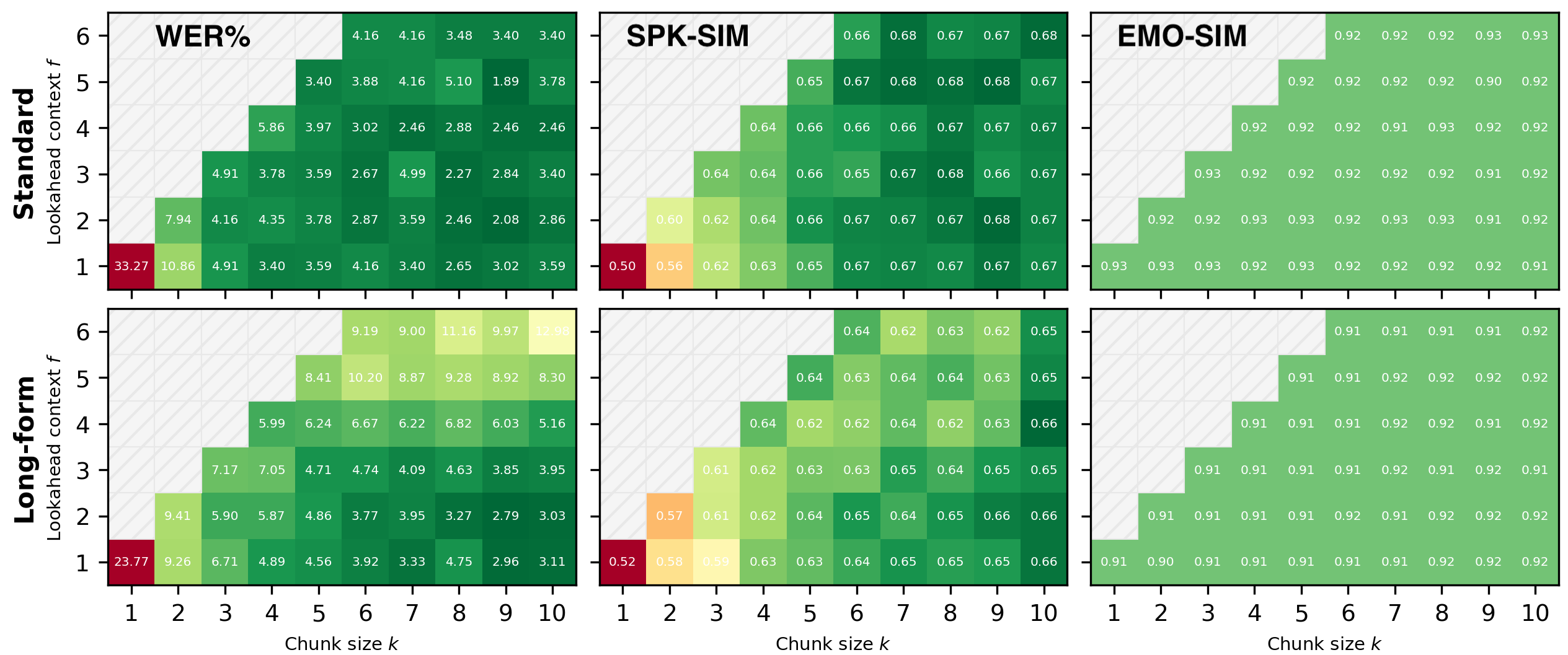}
    \vspace{-0.4cm}
    \caption{Ablation study on quality evaluation with different combinations of chunk size k and lookahead context f.}
    \label{fig:ablation}
\end{figure*}

\textbf{Sliding-Window Baseline}: We implement a simplified sliding-window prompting strategy in which each chunk is conditioned on the previous chunk's generated tokens. Unlike our method, it does not use boundary markers or lookahead context, meaning generation relies solely on past history. Offline batch vocoding is applied to synthesize each chunk.

\textbf{Proposed Method (Boundary-Aware)}: Our method combines sliding-window continuation with the proposed prosodic-boundary marker and a lookahead window of $f=2$ words. Streaming vocoding is employed for incremental audio synthesis. By incorporating limited future context while keeping the effective context length bounded by $\mathcal{O}(k+f)$, the system achieves stable long-form streaming generation.

\section{Results and Discussion}

\subsection{Streaming Latency and Efficiency}

The streaming latency and efficiency results are summarized in Table~\ref{tab:latency}. Our proposed method achieves the lowest TTFA (1296 ms), outperforming both the Interleaved and Sliding-Window baselines. This improvement stems from the boundary-aware prompting mechanism, which enables earlier audio emission than native streaming implementations. For RTF, the Sliding-Window baseline achieves the lowest value (0.718). However, as discussed in Section~\ref{Baselines}, this result is attributed to its use of offline batch vocoding, which maximizes GPU throughput but does not support incremental emission. Between systems using streaming vocoding, our method achieves a lower RTF (0.782) than the Interleaved baseline (0.843). This indicates that bounding the context length improves computational efficiency by preventing unbounded KV-cache growth.

\subsection{Synthesis Quality and Linguistic Fidelity}
Table~\ref{tab:quality} and~\ref{tab:mos} reports the objective and subjective results across both evaluation tiers. Our proposed method consistently achieves the best performance across all metrics.

\subsubsection{Objective Evaluation}
On the \textbf{standard-form} set, our method achieves a WER of 4.03\%, outperforming Interleaved (7.48\%) and Sliding-Window (6.03\%). This shows that the boundary marker and lookahead context stabilize linguistic generation even in short-form synthesis. The model also achieves the highest SPK-SIM (0.64) and EMO-SIM (0.918), suggesting that future context improves speaker consistency and emotional expression.

The performance gap becomes more pronounced in the \textbf{long-form} evaluation. The Interleaved baseline exhibits catastrophic failure, with WER increasing to 70.97\%. This behavior results from \emph{unbounded context growth}: as the KV cache accumulates, the model experiences semantic drift and hallucinations, leading to "garbled" speech and premature termination with severe deletion errors. The Sliding-Window baseline maintains a stable WER (7.83\%) but suffers a severe degradation in speaker similarity (SPK-SIM drops from 0.57 to 0.22). Without boundary markers or lookahead context, the model cannot properly predict sentence boundaries, causing \emph{prosodic drifts} across segments and a decline in speaker and emotional consistency (EMO-SIM drops to 0.857).

In contrast, our Boundary-Aware method maintains stable performance across long-form synthesis, achieving WER 4.77\%, SPK-SIM 0.65, and EMO-SIM 0.912. By bounding the context length and incorporating the lookahead window for prosodic conditioning, the approach effectively mitigates both the hallucination problem of the Interleaved architecture and the prosodic drift of naive Sliding-Window methods.

\subsubsection{Subjective Evaluation}
Subjective listening results confirm the objective findings. While the Interleaved baseline maintains acceptable quality in the standard set (3.99$\pm$0.16), its performance plummet to 3.18$\pm$0.23 in long-form scenario due to linguistic instability. The Sliding-Window baseline degrades further, reaching a MOS of 1.60$\pm$0.18, as listeners frequently observed prosodic discontinuities between segments.

Our Boundary-Aware method achieves the highest perceptual scores across all metrics. In particular, it maintains strong speaker identity (4.24$\pm$0.13 SMOS) and emotional consistency (4.19$\pm$0.13 EMOS) even in long-form synthesis. These results indicate that the proposed boundary-aware conditioning successfully preserves prosodic continuity while maintaining robust streaming generation.

\subsection{Ablation Studies}

As shown in Figure \ref{fig:ablation}, we analyze the trade-off between streaming latency and synthesis quality by varying chunk size $k \in \{1,\ldots,10\}$ and lookahead context $f \in \{1,\ldots,6\}$ under the constraint $f \le k$. For evaluation, we sample a fixed amount of 50 utterances each from both Standard and Long-form tiers. 

Linguistic fidelity (WER) is highly sensitive to the initial context size. At $k=1, f=1$, the model lacks sufficient semantic grounding, resulting in peak WER values of $33.27\%$ and $23.77\%$ for the Standard and Long-form sets. As $k$ increases, performance stabilizes rapidly; for $k \ge 3$, the Standard-tier WER falls below $5\%$. A similar trend is observed in the Long-form tier, indicating that a relatively small semantic anchor is sufficient to stabilize generation.

However, excessive lookahead relative to the chunk size can degrade performance. For instance, at $k=10$, increasing $f$ from $2$ to $6$ raises the Long-form WER from $3.03\%$ to $12.98\%$. This suggests overly strong future conditioning may destabilize generation when it dominates the current segment context.

Speaker and emotion consistency (SPK-SIM and EMO-SIM), are more robust but still benefit from moderate context. Speaker similarity increases from approximately 0.50 at minimal context to above 0.65 for $k \ge 5$, while emotional similarity remains consistently high ($>0.90$) across configurations.

\section{Conclusion}

This paper presents a boundary-aware post-training strategy for streaming LLM-based text-to-speech with streaming text input. By introducing a prosodic-boundary marker and bounded sliding-window prompting, the method stabilizes prosodic generation while preventing unbounded context growth. Experiments on Seed-TTS-Eval show improved streaming latency and synthesis quality over interleaved and sliding-window baselines, while maintaining stable long-form generation with consistent speaker identity and emotion.

Future work will explore generalization to other LLM-based TTS architectures and multilingual settings, as well as adaptive boundary prediction for more flexible streaming generation.




\section{Generative AI Use Disclosure}

During the preparation of this manuscript, the authors used generative AI tools to assist with language refinement, clarity improvement, and \LaTeX{} formatting. These tools were employed solely for editorial support. All scientific content, experimental results, and interpretations were developed and verified by the authors.

\bibliographystyle{IEEEtran}
\bibliography{mybib}

\end{document}